\documentclass[referee]{raa}
\usepackage{graphicx,times}
\usepackage{natbib}
\usepackage{amssymb,amsmath}
 \usepackage[T1]{fontenc}
  \usepackage{aecompl}
\bibpunct{(}{)}{;}{a}{}{,}

\begin{document}

   \title{Photometric Redshift Estimation for Quasars by Integration of KNN and SVM
$^*$ \footnotetext{\small $*$ Supported by the National Natural
Science Foundation of China.} }

 \volnopage{ {\bf 2012} Vol.\ {\bf X} No. {\bf XX}, 000--000}
   \setcounter{page}{1}

   \author{Bo Han\inst{1}, Hongpeng Ding\inst{1}, Yanxia Zhang\inst{2}, Yongheng Zhao\inst{2}}
   \institute{ International School of Software, Wuhan University, Wuhan, 430072,
   P.R.China; \\
        \and Key Laboratory of Optical Astronomy, National Astronomical Observatories,
Chinese Academy of Sciences, 20A Datun Road, Chaoyang District,
100012, Beijing, P.R.China {\it zyx@bao.ac.cn}\\
\vs \no
   {\small Received 2015 June 12; accepted }
}

\abstract{The massive photometric data collected from multiple
large-scale sky surveys offer significant opportunities for
measuring distances of celestial objects by photometric redshifts.
However, catastrophic failure is still an unsolved problem for a
long time and exists in the current photometric redshift estimation
approaches (such as $k$-nearest-neighbor). In this paper, we propose
a novel two-stage approach by integration of $k$-nearest-neighbor
(KNN) and support vector machine (SVM) methods together. In the
first stage, we apply KNN algorithm on photometric data and estimate
their corresponding z$_{\rm phot}$. By analysis, we find two dense
regions with catastrophic failure, one in the range of z$_{\rm
phot}\in[0.3,1.2]$, the other in the range of z$_{\rm phot}\in
[1.2,2.1]$. In the second stage, we map the photometric input
pattern of points falling into the two ranges from original
attribute space into a high dimensional feature space by Gaussian
kernel function in SVM. In the high dimensional feature space, many
outlier points resulting from catastrophic failure by simple
Euclidean distance computation in KNN can be identified by a
classification hyperplane of SVM and further be corrected.
Experimental results based on the SDSS (the Sloan Digital Sky
Survey) quasar data show that the two-stage fusion approach can
significantly mitigate catastrophic failure and improve the
estimation accuracy of photometric redshifts of quasars. The
percents in different |$\Delta$z| ranges and rms (root mean square)
error by the integrated method are $83.47\%$, $89.83\%$, $90.90\%$ and 0.192, respectively,
compared to the results by KNN ($71.96\%$, $83.78\%$, $89.73\%$ and 0.204).
}

   \authorrunning{B. Han et al. }            
   \titlerunning{Photometric Redshift Estimation}  
   \maketitle

\section{Introduction}
Photometric redshifts are obtained by images or photometry. Compared
to spectroscopic redshifts, they show the advantages of high
efficiency and low cost. Especially, with the running of multiple
ongoing multiband photometric surveys, such as SDSS (the Sloan
Digital Sky Survey), UKIDSS (the UKIRT Infrared Deep Sky Survey) and
WISE (the Wide-Field Infrared Survey Explorer), a huge volume of
photometric data are collected, which are larger than spectroscopic
data by two or three orders of magnitude. The massive photometric
data offer significant opportunities for measuring distances of
celestial objects by photometric redshifts. However, photometric
redshifts show the disadvantages of low accuracy compared to
spectroscopic redshifts, and require more sophisticated estimation
algorithms to overcome the problem. Researchers worldwide have
investigated the photometric redshift estimation techniques in
recent years. Basically, these techniques are categorized into two
types: template-fitting models and data mining approaches.
Template-fitting model is the traditional approach for estimating
photometric redshifts in astronomy. It extracts features from
celestial observational information, such as multiband values, and
then matches them with the designed templates constructed by
theoretical models or real observations. With feature matching,
researchers can estimate photometric redshifts. For example,
Bolzonella et~al. (2000) estimated photometric redshifts through a
standard SED fitting procedure, where SEDs (spectral energy
distributions) were obtained from broad-band photometry. Wu et~al.
(2004) estimated the photometric redshifts of a large sample of
quasars with the $\chi^2$ minimization technique by using derived
theoretical color-redshift relation templates. Rowan-Robinson et~al.
(2008) proposed an approach using fixed galaxy and quasar templates
applied to data at 0.36-4.5 $\mu m$, and on a set of four infrared
emission templates fitted to infrared excess data at 3.6-170$\mu m$.
Ilbert et~al. (2009) applied a template-fitting method (Le Phare) to
calculate photometric redshifts in the 2-deg$^2$ COSMOS field.
Experimental results from the above template-fitting methods showed
that their estimation accuracy relied on the templates constructed
by either simulation or real observational data.

Data mining approaches apply statistics and machine learning
algorithms on a set of training samples and automatically learn
complicated functional correlations between multiband photometric
observations and their corresponding high confidence redshift
parameters. These algorithms are data-driven approaches, rather than
template-driven approaches. The experimental results showed that
they achieved much accurate photometric estimations in many
applications. For example, Ball et~al. (2008) applied a nearest
neighbor algorithm to estimate photometric redshifts for galaxies
and quasars using SDSS and GALEX (the Galaxy Evolution Explorer)
data sets. Abdalla et~al. (2008) estimated photometric redshifts by
using a neural network method. Freeman et~al. (2009) proposed a
non-linear spectral connectivity analysis for transforming
photometric colors to a simpler, more natural coordinate system
wherein they applied regression to make redshift estimations. Gerdes
et~al. (2010) developed a boosted decision tree method, called
ArborZ, to estimate photometric redshifts for galaxies. Way et~al.
(2012) proposed an approach based on Self-Organizing-Mapping (SOM)
to estimate photometric redshifts. Bovy et~al. (2012) presented the
extreme deconvolution technique for simultaneous classification and
redshift estimation of quasars and demonstrated that the addition of
information from UV and NIR bands was of great importance to
photometric quasar-star separation and essentially the redshift
degeneracies for quasars were resolved. Carrasco et~al. (2013)
presented an algorithm using prediction trees and random forest
techniques for estimating photometric redshifts, incorporating
measurement errors into the calculation while also efficiently
dealing with missing values in the photometric data. Brescia et~al.
(2013) applied the Multi Layer Perceptron with Quasi Newton
Algorithm (MLPQNA) to evaluate photometric redshifts of quasars with
the data set from four different surveys (SDSS, GALEX, UKIDSS, and
WISE).

Though template-fitting approaches and data mining approaches can
roughly estimate photometric redshifts, they both suffer the
catastrophic failure problem in estimating photometric redshifts of
quasars when the spectroscopic redshift is less than 3 (Richards et~al. 2001; Weinstein
et~al. 2004; Wu et~al. 2004). Zhang et~al.
(2013) practically demonstrated that with cross-matched multiband
data from multiple surveys, such as SDSS, UKIDSS and WISE, $k$-nearest
neighbor (KNN) algorithm can largely solve the catastrophic failure problem
and improve photometric redshift estimation accuracy. The method
becomes more important as the development of multiple large
photometric sky surveys and the coming of the age of astronomical
big data. However, during the data preparation process, we need to cross-match
multiband information of quasars from multiple photometric surveys.
The number of matched quasar records is far less than the original
quasar number in a single survey. For example, there are 105,783
quasar samples available in SDSS DR7. However, the number of
cross-matched samples from SDSS, WISE and UKIDSS is only 24,089. The
cross-matched sample is around one fourth of SDSS quasar data. This
shortcoming greatly limits the application scope of this
estimation approach to only a small portion of
cross-matched quasars observed by all surveys.

In this paper, we propose a novel two-stage photometric redshift
estimation approach, i.e., the integration of KNN ($k$-nearest
neighbor) and SVM (support vector machine) approaches, to mitigate
catastrophic failure for quasars by using relative few band
attributes only from a single survey. The paper is organized as
follows. Section 2 describes the data used. Section 3 presents a
brief overview of KNN, SVM and KNN+SVM. Section 4 gives the
experimental results by KNN+SVM. The conclusions and discussions are
summarized in Section 5.

\section{Data}

Our experiments are based on a dataset generated from the Sloan
Digital Sky Survey (SDSS; York et~al. 2000), which labels highly
reliable spectroscopic redshifts and has been widely used in
photometric redshift estimation. The dataset was constructed by
Zhang et~al. (2013) for estimating photometric redshifts of quasars.
They used the samples of the Quasar Catalogue V (Schneider et~al.
2010) in SDSS DR7, which included 105,783 spectrally confirmed
quasars. In each quasar record, five band features
$u,g,r,i,z$ are provided. Similar to Zhang et~al. (2013), in our experiments, we use
these five attributes $u-g,g-r,r-i,i-z,r$ (short for $4C+r$) as the input and the corresponding
spectroscopic redshift as a regression output.

\section{Methodology}

Firstly, we study the characteristics of catastrophic failure for
quasars and observe that the outlier points by KNN are clustered
into two groups: one group's spectroscopic redshift z$_{\rm spec}$ is between 0.2
and 1.1, while its photometric redshift z$_{\rm phot}$ is between 1.2 and 2.1, and
the other group's z$_{\rm spec}$ is between 1.4 and 2.3,
while its z$_{\rm phot}$ is between 0.3 and 1.2 (shown in Figure~1). Some points
with z$_{\rm phot}$ falling into Group 1 actually have z$_{\rm spec}$ close to the
range of Group 2, but they are wrongly estimated by KNN and are
mixed into Group 1, and vice versa. The two
groups look almost 180-degree rotationally symmetric along the 45-degree diagonal line in
the z$_{\rm phot}$ vs. z$_{\rm spec}$ diagram.
The two outlier clusters show that KNN method cannot
effectively distinguish outlier points from good estimation
points using Euclidean distance in the two regions. Next, we propose
a two-stage integration approach by fusion of $k$-nearest neighbors
(KNN) and support vector machine (SVM) methods. In the first stage,
we apply KNN algorithm on photometric data and estimate their
corresponding z$_{\rm phot}$. In the second stage, we map
photometric multiband input pattern of points falling into the two
ranges with z$_{\rm phot}\in [0.3,1.2]$ and z$_{\rm phot}\in
[1.2,2.1]$ from an original attribute space into a high dimensional
feature space by Gaussian kernel function in SVM. In the high
dimensional feature space, many outlier points can be
identified by a classification hyperplane in SVM and further be
corrected. Since most points of catastrophic failure have been
identified and corrected, our integration approach can improve the
photometric redshift estimation accuracy.

\begin{figure}
   \centering
   \includegraphics[width=12cm,clip]{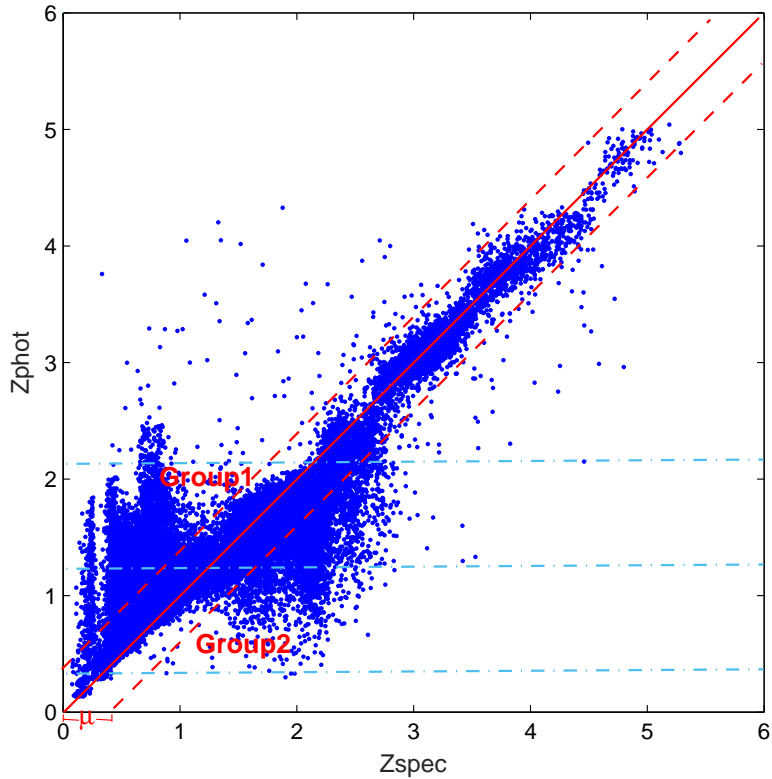}
   \caption
   { \label{fig:example}
Photometric redshift estimation by KNN estimation. The points in Group 1 and Group 2 are outlier points. $\mu$ is the parameter representing the error tolerant scope. The slant dotted lines give the corrected estimated range of photometric redshifts. The horizontal dotted lines describes the zones of Group 1 and Group 2.}
   \end{figure}

The KNN algorithm generally applies Euclidean distance of attributes
(shown in Equation~1) to compute distance between point $m$ and
point $n$,
\begin{equation}
d_{m,n}=\sqrt{(f_{m,1}-f_{n,1})^2+(f_{m,2}-f_{n,2})^2+...+(f_{m,k}-f_{n,k})^2}
\end{equation}
where $f_{m,j}$($f_{n,j}$) denotes the $j$th attribute among $4C+r$
input pattern for the $m$th ($n$th)
points, $k$ represents the total number of attributes. The points in
Group 1 and Group 2 show that those outlier quasars cannot be
correctly identified in an Euclidean space. In other words, we cannot
have a simple plane as a useful separating criterion between points
in Group 1 and Group 2. Based on the present data, the provided
information is not enough to give good estimation of the outlier
points. Now there is a question whether those outlier points can be
linearly separable in a high-dimensional non-Euclidean feature
space? Thereby, we explore the kernel function in SVM and map
features into a high dimension space and test if we can correctly
classify outlier points in Group 1 and Group 2. From the analysis
above, we propose a two-stage integration approach by fusion of
estimation with KNN and classification with SVM.

\subsection{Estimation with K-Nearest-Neighbor}

K-Nearest-Neighbor (KNN) algorithm is a lazy predictor which
requires a training set for learning. It first finds the nearest
neighbors by comparing distances between a test sample and training
samples that are similar to it in a feature space. Next, it assigns
the average value of the nearest neighbors to the test sample as its
prediction value. In general, the distance is computed as
Euclidean distance described in Equation~1. In the era of big data,
we have been collecting more data than ever before and KNN achieves
much accurate predictions (Zhang et~al. 2013). Thereby, we also use
KNN in our research. One disadvantage of KNN is the high computation
cost. We apply KD-tree to efficiently implement KNN algorithm.

\subsection{Classification with Support Vector Machine}

Support Vector Machine (SVM) is an effective classification
algorithm based on structural risk minimization principle proposed
by Vapnik (1995). Given a training dataset with $n$ records,
each record has the pattern of ($x_i,y_i$) for $i=1,2,...,n$, we
aim to build a linear classifier with the following Equation~2,
\begin{equation}
f(x)=w\cdot x+b
\end{equation}
Here, $w$ and $b$ are weight vector and bias respectively. Figure~2
illustrates that several lines can separate two categories of
points. In SVM, for minimizing the classification error risk for
other test datasets, we aim to find a line (shown as the dash line
in Figure~2) with the maximized margin to separate the two classes
of points. This principle makes SVM have a better classification
accuracy than other competing machine learning models in many
classification tasks.

\begin{figure}
   \centering
      \includegraphics[bb=5 0 329 233,width=12cm,clip]{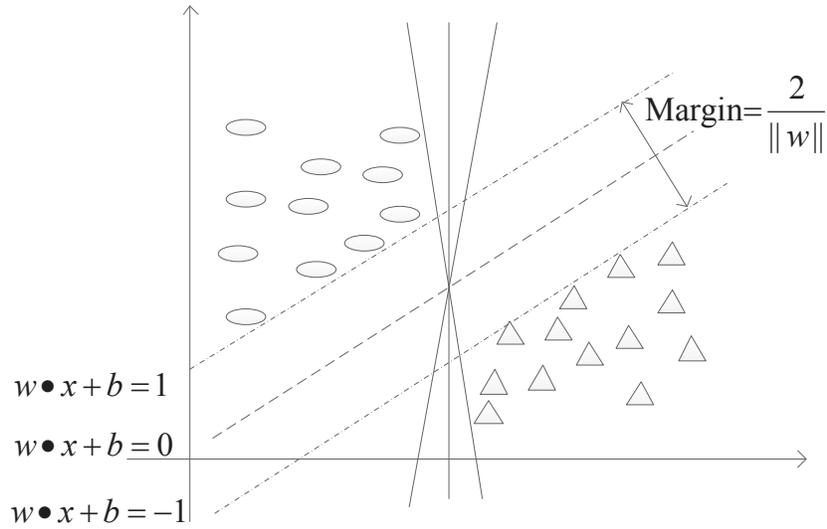}
      \caption
   { \label{fig:example}
Maximizing classification margin is aimed in SVM. The points on the
dotted lines are called as support vectors. The distance between the
two dotted lines is named as Margin. When the Margin is maximized, the classification accuracy achieves best.}
   \end{figure}

\begin{figure}
    \centering
   \includegraphics[bb=1 2 485 197,width=12cm,clip]{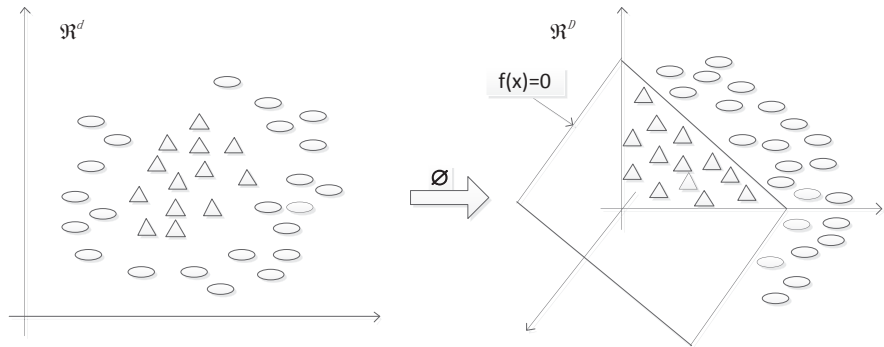}
   \caption
   { \label{fig:example}
Linear indistinguishable points in a low dimensional space ($\Re^d$) can be
separated in a high dimensional space ($\Re^D$) by the application of kernel
function ($\emptyset$) in SVM. $f(x)=0$ represents the hyperplane that separates the two classes.}
   \end{figure}

Sometimes, a classification task is hard and not linearly solvable.
The left graph in Figure~3 shows one such case. In such case, by
Vapnik-Chervonenk in dimension theory, SVM applies a kernel function
to promote the original flat space in the ordinary inner product
concepts. By the theory of reproducing kernels, we can map the
original Euclidean feature space to the high-dimensional
non-Euclidean feature space in SVM classification algorithm.
Thereby, some of non-linearity problems in the original
low-dimensional feature space $\Re^d$ become linearly solvable in
the high-dimensional space $\Re^D$. The right graph in Figure~3
shows a dimension mapping by kernel function to solve the problem.
Therefore, Equation~2 can be transformed to the following form by
feature mapping function $\emptyset$,

\begin{equation}
f(x)=w\cdot \emptyset(x)+b
\end{equation}

In this way, we can have the following objective function and
constraints for a SVM classifier as below, and minimize $$\mid\mid
w\mid\mid^2+C\sum_{i=1}^{n}\epsilon_i
$$
subject to
\begin{equation}
y_i\cdot(<w,\emptyset(x_i)>+b)\ge 1-\epsilon_i
\end{equation}
here, $C$ is a regularization parameter and $\epsilon_i$ is a slack
variable.

By represented theorem, we have,
\begin{equation}
f(x)=\sum_{i=1}^{n}\alpha_iy_i\emptyset(x_i)^T\emptyset(x)+b
\end{equation}

$\alpha_i$ is a parameter with the constraint that $\alpha_i \ge 0$.
For solving Equation~5, SVM introduces a kernel function defined
as,
\begin{equation}
K(x_i,x)=\emptyset(x_i)^T\emptyset(x)
\end{equation}

In this paper, we practically apply Gaussian kernel (shown in
Equation~7) to achieve the non-linear classification.
\begin{equation}
K(x_i,x)=e^{-\frac{\mid\mid x_1-x_2\mid\mid^2}{2\sigma^2}}
\end{equation}
where $x_1, x_2$ represents vectors of multiband attributes or
input patterns observed from a single survey, $\sigma$ is a free
parameter.

In this way, we aim to apply a SVM classifier to distinguish the
mixture points in Group 1 and other points around the minor diagonal
with z$_{\rm phot}\in[1.2,2.1]$ in the z$_{\rm phot}$ vs. z$_{\rm
spec}$ diagram. Similarly, we can distinguish points in Group 2 and
other points with z$_{\rm phot}\in[0.3,1.2]$.

\subsection{Integration of KNN and SVM for Photometric Redshift Estimation}

The photometric redshift estimation algorithm by integration of KNN
and SVM is presented in the following. To obtain the robust accuracy
measure for our integration approach, we repeat the experiments for
$num$ rounds. In each round, the data sets will experience the
initialization step, KNN step, SVM training step, SVM test step,
correction step and evaluation step. In initialization step, we
randomly divide the SDSS data set into separate training set, validation set
and test set. In KNN step, we apply KNN
algorithm ($k=17$) to estimate z$_{\rm phot-validation}$ and z$_{\rm
phot-test}$ based on training set and the union of training set and
validation set respectively. In SVM training step, we aim to build
two SVM classifiers: SVM1 and SVM2 to distinguish good estimation and
outlier points with z$_{\rm phot-validation}\in[1.2,2.1]$ and
z$_{\rm phot-validation}\in[0.3,1.2]$, respectively. The good
estimation or outlier points is defined in the following Equation~8,
\begin{equation}
\left\{
\begin{array}{rcl}
\mid z_{\rm spec}-z_{\rm phot}\mid &\le&\mu\,\,\,\,\, {\rm good}\\
\mid z_{\rm spec}-z_{\rm phot}\mid &>&\mu\,\,\,\,\, {\rm bad}
\end{array}
\right.
\end{equation}
here, $\mu$ is the parameter which means the error tolerant scope derived from the validation set.

Visually, the good estimation points will fall into area close to
45-degree diagonal line in the diagram, while the outlier
points will fall into Group 1 and Group 2 in Figure~1.

Specifically, we use those outlier points with z$_{\rm
phot-validation}\in[1.2,2.1]$ and z$_{\rm phot-validation}\in
[0.3,1.2]$ to construct datasets: Group1\_trainingdata and
Group2\_trainingdata, respectively. In the two datasets, inputs are
patterns $4C+r$ and z$_{\rm phot}$ directly from KNN, and the output is z$_{\rm spec}$.

In SVM test step, we apply classifiers SVM1 and SVM2 to identify
outlier points.

In correction step, we use KNN algorithm based on Group1\_data to
compute z$_{\rm phot}$ for those outlier points distinguished by
SVM1 in test data. Since Group1\_data and those outlier points have
the similar pattern while the output of Group1\_trainingdata is
z$_{\rm spec}$, the KNN algorithm can improve the z$_{\rm phot}$
estimation. Similarly, we can use Group 2 to train data and then correct
outlier points distinguished by SVM2 in test data.

In evaluation step, we apply the percents in different |$\Delta$z| ranges and root mean square (rms)
error of $\Delta$z to test our photometric redshift estimation
approach. The definition of $\Delta$z is listed in
Equation~9.
\begin{equation}
\Delta z=\frac{z_{\rm spec}-z_{\rm phot}}{1+z_{\rm spec}}
\end{equation}

The detailed steps of the two-stage method are as following. To be
much clearer, the flow chart of the whole process is shown in
Figure~4.

LoopId$=1$;

Do while LoopId$\le num$;

Initialization Step:

\quad Randomly select 1/3 sample from the SDSS quasar sample as the training set,
another 1/3 sample as the validation set, and the remaining 1/3
sample as the test set.

KNN Step:

\quad 1. Based on training set, we apply KNN ($k=17$) algorithm to
estimate z$_{\rm phot-validation}$ for each sample in validation
set;

\quad 2. Based on the union of training set and validation set, we
apply KNN ($k=17$) algorithm to estimate z$_{\rm phot-test}$ for each sample
in test set.

SVM Training Step:

\quad 1. For those samples with z$_{\rm phot-validation}\in
[1.2,2.1]$ in validation set, we train a classifier SVM1 with
Gaussian kernel, which distinguishes good estimation and outlier
points by Equation~8. With those outlier points, we build a data set
Group1\_trainingdata, which is composed of $4C+r$ and z$_{\rm phot}$ as the input and
z$_{\rm spec}$ as the output;

\quad 2. Similarly, for those samples with z$_{\rm
phot-validation}\in [0.3,1.2]$ in validation set, we train a classifier SVM2 with Gaussian kernel, which distinguishes good estimation and outlier points. With those outlier points, we build a
data set Group2\_trainingdata, which is composed of $4C+r$ and z$_{\rm phot}$ as the
input and z$_{\rm spec}$ as the output.

SVM Test Step:

\quad 1. For those samples with z$_{\rm phot-test}\in [1.2,2.1]$ in
test set, we apply the classifier SVM1 to distinguish good
estimation and outlier points;

\quad 2. Similarly, for those samples with z$_{\rm phot-test}\in
[0.3,1.2]$ in test set, we apply the classifier SVM2 to
distinguish good estimation and outlier points.

Correction Step:

\quad 1. For those outlier points with z$_{\rm phot-test}\in
[1.2,2.1]$ in test set, we apply the KNN algorithm based on the data
set Group1\_trainingdata;

\quad 2. For those outlier points with z$_{\rm phot-test}\in
[0.3,1.2]$ in test set, we apply the KNN algorithm based on the data
set Group2\_trainingdata.

Evaluation Step:

\quad By comparing z$_{\rm phot-test}$ and z$_{\rm spec}$ for all
samples in test set, we compute the popular accuracy measures for
redshift estimation and rms error of $\Delta$z.

LoopId=LoopId+1;

End do.

Output the mean and standard error for the percents in different |$\Delta$z| ranges and rms error of $\Delta$z to evaluate the accuracy of our proposed integrated approach KNN+SVM.

\begin{figure}[!!!h]
   \centering
   \includegraphics[width=16cm,clip]{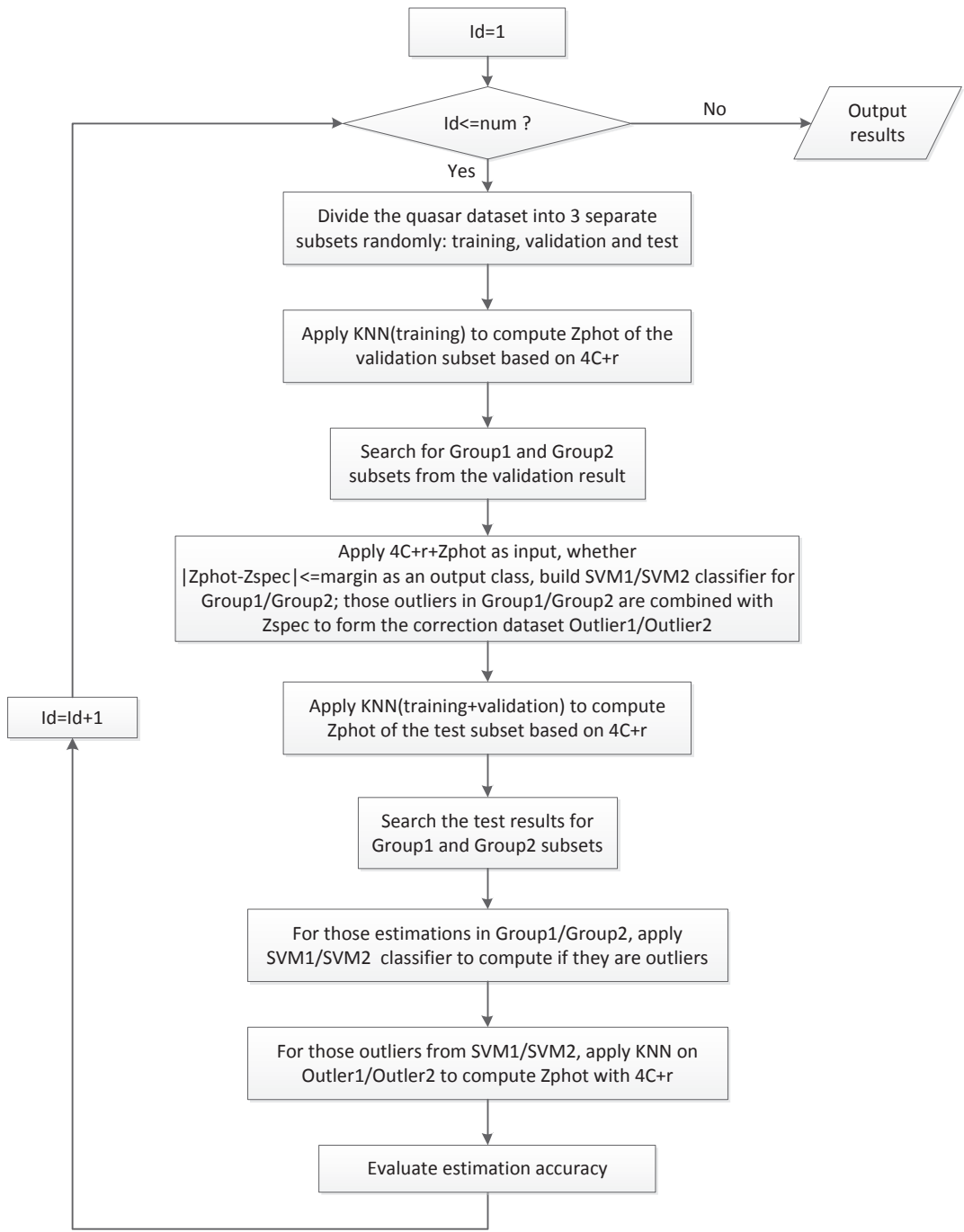}
   \caption
   { \label{fig:example}
The flow chart of photometric redshift estimation by the integration
of KNN and SVM.}
   \end{figure}

\section{Experimental results}

In the experiments, we adopt the input pattern $4C+r$ as attributes,
which are widely accepted by recent researches on photometric
redshift estimation. In our designed algorithm, we practically set
$num=10$ and repeat the experiments for 10 times.

For classification, we apply the widely used tool LIBSVM (Chang,
2011). By using Gaussian kernel function, we train classifiers
SVM1 and SVM2 for the sample with z$_{\rm phot-validation}\in
[1.2,2.1]$ and for the sample with z$_{\rm phot-validation}\in
[0.3,1.2]$ in the validation set, separately. To optimize the estimation
accuracy, we adjust two parameters controlling the
Gaussian kernel in SVM, a cost coefficient $C$ that measures the
data unbalance and a factor $\gamma$ depicting the shape of the high
dimensional feature space. Other parameters are set to the default
values in LIBSVM. In order to obtain the best model parameters, the
grid search is adopted. The grid search in SVM1 and SVM2 is
indicated in Figure~5. For SVM1, the optimal model parameter $C$ is
2, $\gamma$ is 8, meanwhile, the classification accuracy is 94.12\%.
For SVM2, the best model parameter $C$ is 128, $\gamma$ is 0.5, the
classification accuracy amounts to 90.04\%.

\begin{figure}
   \centering
   \includegraphics[width=16cm,clip]{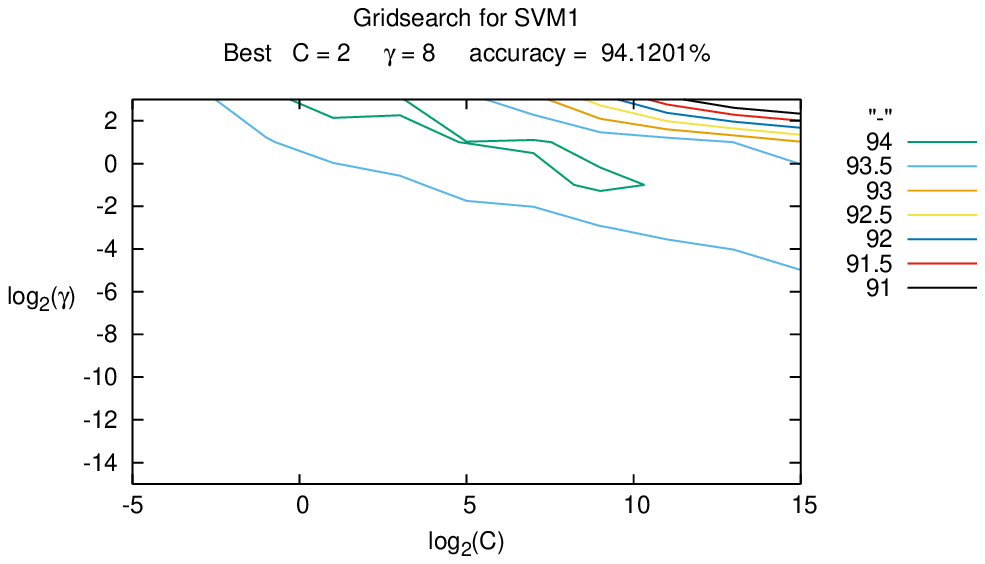}
   \includegraphics[width=16cm,clip]{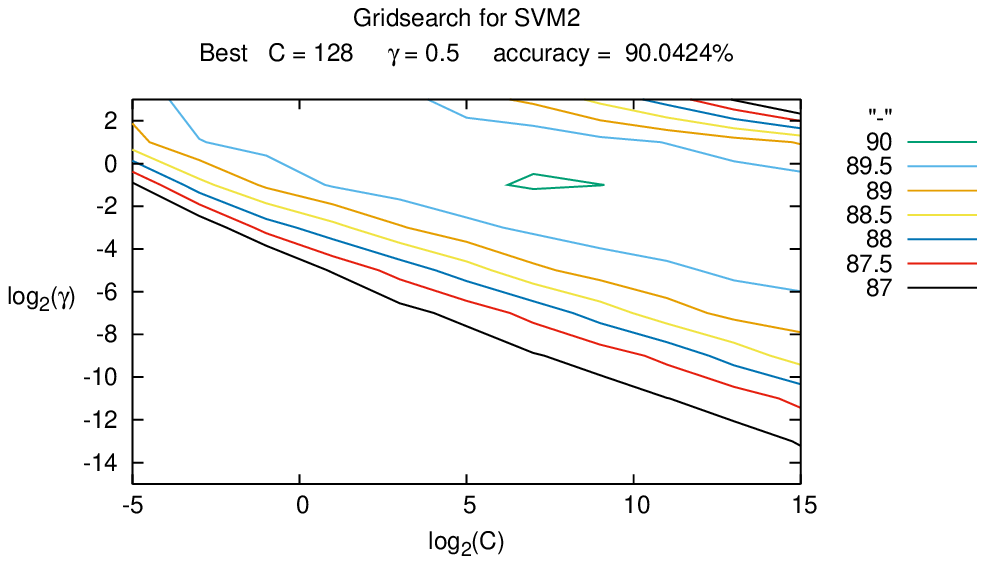}
   \caption
   { \label{fig:example}
Top: the best model parameter in SVM1 is obtained by grid search,
i.e. $C$=2, $\gamma$=8, the accuracy of classification achieves
94.12\%. Bottom: the best model parameter in SVM2 is obtained by grid
search, i.e. $C$=128, $\gamma$=0.5, the accuracy of classification
is 90.04\%.}
   \end{figure}

With the optimized parameters and the union of the training set and
the validation set as a new training set, we compare the estimation
accuracy between original KNN ($k=17$) algorithm and our integration
approach KNN+SVM. The parameter $\mu$ is a factor to
determine whether a point has good estimation or not. We change the
value of $\mu$ to check its influence on the estimation accuracy.
The results are listed in Table 1. For KNN, the proportions of
|$\Delta$z|$<0.1,0.2,0.3$ and rms error of predicted photometric
redshifts are 71.96\%, 83.78\%, 89.73\% and 0.204, separately; for
KNN+SVM, these optimal measures are 83.47\%, 89.83\%, 90.90\% and
0.192, respectively, when $\mu$=0.3, which are bold in Table~1.
Obviously, these criteria for photometric redshift estimation are
all significantly improved with the new method. It suggests that the
integration approach can effectively correct those outlier points
with z$_{\rm phot}\in[1.2,2.1]$ and z$_{\rm phot}\in[0.3,1.2]$.
Thereby, it can significantly mitigate catastrophic failure and
improve the estimation accuracy of photometric redshifts.

\begin{table*}
\begin{center}
\caption{Comparison of KNN and our integration approach}
\bigskip
\begin{tabular}{lcccl}
\hline\hline
Method&|$\Delta$z|$<0.1(\%)$&|$\Delta$z|$<0.2(\%)$&|$\Delta$z|$<0.3(\%)$& rms error\\
\hline
KNN($k=17$)         & 71.96$\pm$0.20 & 83.78$\pm$0.18 &89.73$\pm$0.16 &0.204$\pm$0.004\\
SVM+KNN($\mu=$0.1)& 75.06$\pm$3.03 & 81.43$\pm$2.31 &85.51$\pm$1.69 &0.232$\pm$0.022\\
SVM+KNN($\mu=$0.2)& 80.86$\pm$1.19 & 85.56$\pm$1.95 &86.57$\pm$1.81 &0.224$\pm$0.013\\
\bf SVM+KNN($\mu=$0.3)&\bf 83.47$\pm$0.86 &\bf 89.83$\pm$0.51 &\bf 90.90$\pm$0.42 &\bf 0.192$\pm$0.007\\
SVM+KNN($\mu=$0.4)& 81.63$\pm$0.64 & 89.53$\pm$0.32 &91.54$\pm$0.33 &0.193$\pm$0.005\\
SVM+KNN($\mu=$0.5)& 78.89$\pm$0.22 & 88.30$\pm$0.24 &91.63$\pm$0.21 &0.194$\pm$0.005\\
SVM+KNN($\mu=$0.6)& 75.84$\pm$0.14 & 86.60$\pm$0.13 &90.58$\pm$0.11 &0.199$\pm$0.003\\
\hline \hline
\end{tabular}
\bigskip
\end{center}
\end{table*}

The experimental results also show that without cross-matching
multiband observations from multiple surveys, we can effectively
apply Gaussian kernel function in SVM to identify outlier points in
Group 1 and Group 2 from catastrophic failure by mapping attributes
from a single data source into a high dimensional feature space. The
identification helps us correct those outlier points and thereby
improve estimation accuracy.

In order to compare the performance of photometric redshift
estimation by KNN algorithm with that by KNN and SVM approach, the
photometric redshift estimation with these two methods is shown in
Figure~6 and Figure~7, respectively. As indicated by Figures 6-7, we
can see clearly that the outlier points in both Group 1 and Group 2
have been significantly decreased by adopting the new method
KNN+SVM. It intuitively proves that our proposed approach is
effective.

\begin{figure}
   \centering
\includegraphics[width=10cm,height=10cm,clip]{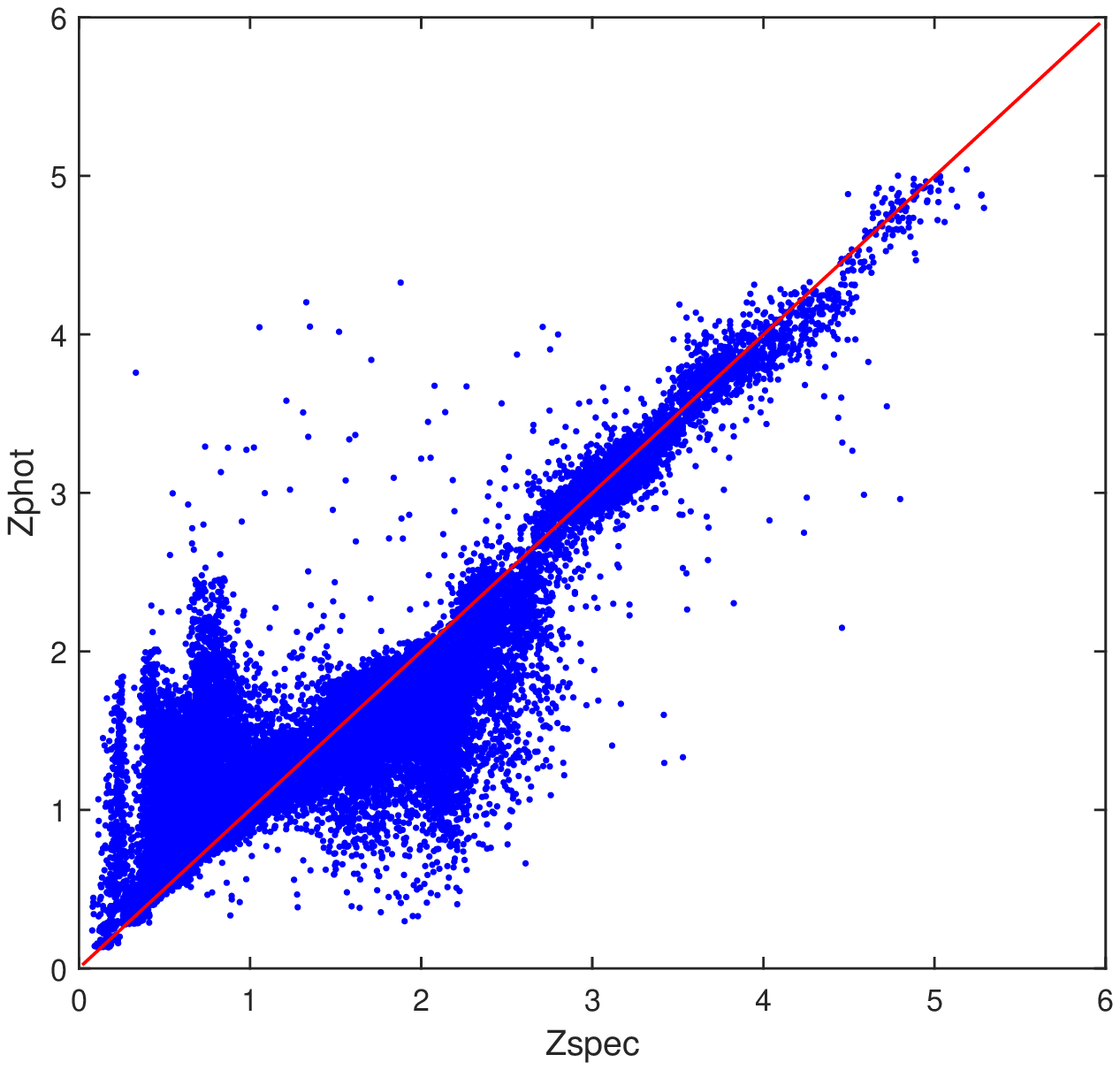}
\caption[fig4] {Photometric redshift estimation by KNN.}
\end{figure}

\begin{figure}
   \centering
\includegraphics[width=10cm,height=10cm,clip]{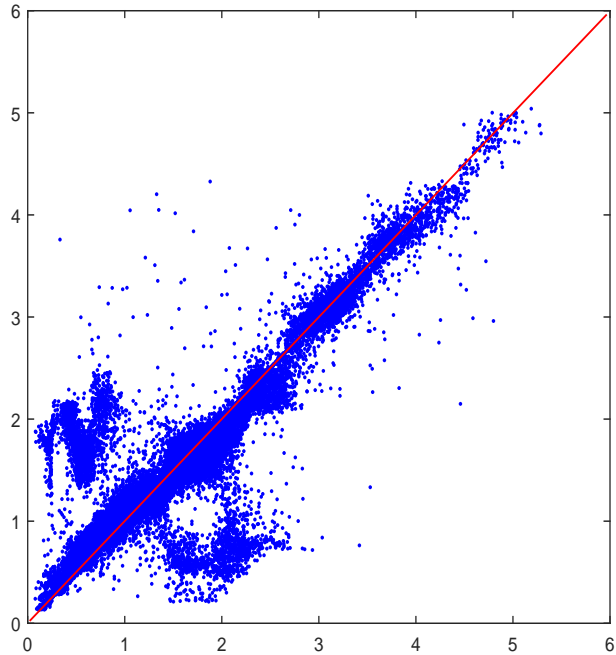}
\caption[fig4] {Photometric redshift estimation by KNN+SVM.}
\end{figure}

\section{Conclusions and Discussions}

Catastrophic failure is an unsolved problem with a long history
existing in most photometric redshift estimation approaches. In this
paper, we firstly analyze the reasons of catastrophic failure for
quasars and point out that the outlier points result from being
non-linearly separable in Euclidean feature space of input pattern.
Next, we propose a new estimation approach by integration of KNN and
SVM methods together. By Gaussian kernel function in SVM, we map
multiband input pattern from an original Euclidean space into a high
dimensional feature space. In this way, many outlier points can be
identified by a hyperplane and then corrected. The experimental
results based on SDSS data for quasars show that the integration
approach can significantly mitigate catastrophic failure and improve
the photometric redshift estimation accuracy, e.g. the
percentages in different |$\Delta$z| ranges and rms error are $83.47\%$, $89.83\%$, $90.90\%$ and 0.192, respectively. While different previous researches of mitigating catastrophic failure by
cross-match of data from several surveys, our approach can achieve
the similar objective only from a single survey and needn't cross-match among multiple surveys avoiding cross-match efforts especially for the growing of large survey data. Moreover, not all sources have observation from different surveys. Therefore this method can be widely applied for a single large sky survey photometric data. In addition, the integration method with data from more bands may
further improve the accuracy of estimating photometric redshifts of
quasars.

\begin{acknowledgements}
We are very grateful to the referee's important comments and
suggestions which help us improve our paper. This work is supported
by the National Natural Science Foundation of China under Grants
NO.61272272 and NO.U1531122, National Key Basic Research Program of China
2014CB845700 and NSFC-Texas A\&M University Joint Research Program
No.11411120219. We acknowledgment SDSS database. The SDSS is managed
by the Astrophysical Research Consortium for the Participating
Institutions. The Participating Institutions are the American Museum
of Natural History, Astrophysical Institute Potsdam, University of
Basel, University of Cambridge, Case Western Reserve University,
University of Chicago, Drexel University, Fermilab, the Institute
for Advanced Study, the Japan Participation Group, Johns Hopkins
University, the Joint Institute for Nuclear Astrophysics, the Kavli
Institute for Particle Astrophysics and Cosmology, the Korean
Scientist Group, the Chinese Academy of Sciences (LAMOST), Los
Alamos National Laboratory, the Max-Planck-Institute for Astronomy
(MPIA), the Max-Planck-Institute for Astrophysics (MPA), New Mexico
State University, Ohio State University, University of Pittsburgh,
University of Portsmouth, Princeton University, the United States
Naval Observatory, and the University of Washington.
\end{acknowledgements}

\end{document}